%% file: main.tex
\documentclass[conference]{IEEEtran}

\usepackage{multicol}

\usepackage{xcolor}
\usepackage{soul}
\usepackage{amsmath}
\usepackage{graphics}
\usepackage[demo]{graphicx}
\usepackage{subfig}
\usepackage{setspace}
\usepackage{cite}
\usepackage{latexsym}
\usepackage{float}
\usepackage{epsfig}
\usepackage{multirow}
\usepackage{cite,cases,url}
\usepackage{amssymb}
\usepackage{graphicx}
\usepackage{epstopdf}
\usepackage{balance}
\usepackage{enumerate}
\usepackage{array}
\usepackage{algorithmic}
\usepackage{algorithm}
\usepackage{comment}
\usepackage[normalem]{ulem}
\useunder{\uline}{\ul}{}

\newcolumntype{L}{>{\centering\arraybackslash}m{5cm}}
\newcolumntype{K}{>{\centering\arraybackslash}m{6cm}}
\newcolumntype{P}{>{\centering\arraybackslash}m{2.3cm}}
\newcolumntype{M}{>{\raggedright\arraybackslash}m{2cm}}
\newcolumntype{N}{>{\raggedright\arraybackslash}m{2.5cm}}

\begin{document}

\title{Securing Mobile IoT with Unmanned Aerial Systems}

\author{\IEEEauthorblockN{Aly Sabri Abdalla\IEEEauthorrefmark{1},
Bodong Shang\IEEEauthorrefmark{2},
Vuk Marojevic\IEEEauthorrefmark{1},
Lingjia Liu\IEEEauthorrefmark{2}\\
}
\IEEEauthorblockA{\IEEEauthorrefmark{1}Department of Electrical and Computer Engineering, Mississippi State University, MS 39762, USA\\}
\IEEEauthorblockA{\IEEEauthorrefmark{2}The Bradley Department of Electrical and Computer Engineering, Virginia Tech, Blacksburg VA, 24061, USA\\}
Email: asa298@msstate.edu, bdshang@vt.edu, vuk.marojevic@msstate.edu, ljliu@vt.edu
}

\maketitle


\begin{abstract}
\input{./include/abs.tex}
\end{abstract}

\IEEEpeerreviewmaketitle

\section{Introduction}
\label{sec:intro}
\input{./include/intro.tex}
\section{Wireless Security Fundamentals}
\label{sec:req}
\input{./include/requirements.tex}

\section{Problem Formulation and Related Work}
\label{sec:probem}
\input{./include/problem.tex}

\section{System Model}
\label{sec:system}
\input{./include/system.tex}


\section{Simulation Results and Discussion}
\label{sec:results}
\input{./include/results.tex}
\section{Conclusions}
\label{sec:conclusions}
\input{./include/conclusions.tex}

\section*{Acknowledgement}
The work of A. S. Abdalla and V. Marojevic was supported in part by the NSF Platforms for Advanced Wireless Research (PAWR) program under grant number CNS-1939334, the Bagley College of Engineering and the Department of Electrical and Computer Engineering at Mississippi State University.



\balance

\bibliographystyle{IEEEtran}
\bibliography{main}


\end{document}

%% file: include/abs.tex
The Internet of Things (IoT) will soon be omnipresent and billions of sensors and actuators will support our industries and well-being. 
IoT devices are embedded systems that are connected using wireless technology for most of the  cases. 
The availability of the wireless network serving the IoT, the privacy, integrity and trustworthiness of the data are of critical importance, since IoT will drive businesses and personal decisions.
This paper proposes a new approach 
in the wireless security domain that leverages advanced wireless technology and the emergence of the unmanned aerial system or vehicle (UAS or UAV). 
We consider the problem of eavesdropping and analyze how UAVs can aid in reducing, or overcoming this threat in the mobile IoT context. 
The results show that huge improvements in terms of channel secrecy rate can be achieved when UAVs assist base stations for relaying the information to the desired IoT nodes. 
Our approach is technology agnostic and can be expanded to address other communications security aspects.

\textit{Index Terms}--IoT, wireless communications, secrecy rate, security, UAS, UAV.

%% file: include/intro.tex

Wireless communications networks are omnipresent and provide nationwide coverage for commercial use in many countries.
Mobile users or devices today are mainly served by cellular communications services. 
The 4G long-term evolution (LTE) is widespread and provides broadband mobile access to terrestrial users.
New use cases and applications for advanced wireless  systems, especially in the industry automation and vehicular control, require more stringent performance features in terms of reliability, latency and security. 
The 5G ecosystem will provide an overarching framework for delivering flexible and customizable networking and end-to-end services. 
It will connect simple sensors, actuators, user devices, sophisticated industrial control systems, medical systems, vehicles, cities, and critical infrastructure~\cite{8519960}. 
It will enable massive Internet of Things (IoT) deployments, where low cost, low power, and low complexity are critical~\cite{8642801}. 

In an era where digital data dominates businesses and society as a whole, the IoT has emerged as a natural evolution of embedded systems that have been supporting many industries and end users for decades. 
Billions of IoT devices will soon be deployed for supporting precision agriculture, search and rescue, smart transportation, critical infrastructure, and so forth. 
What all these devices have in common is that they need to be connected to the Internet. 
Wireless communications is the only viable solution for most of the IoT use cases because of practical deployment considerations \cite{6730660}. 

Cellular communications networks will play a major role for providing  connectivity to the IoT \cite{8383979}. The Third Generation Partnership Project (3GPP) therefore introduced the narrowband IoT (NB-IoT) standard as part of its LTE framework in Release 13. It allows using LTE systems to serve IoT devices using the smallest allocateable resource, one resource block~\cite{wang2017primer}.

Because of the broad IoT use cases, from user-centric applications to business-centric data, security becomes a very important element of the IoT~\cite{8688434}. 
But, security is often not rigorously implemented for most of the cases because of resource limitations, among others. 
The focus of this paper is on secure communications systems that serve the IoT. 
Security of communications encompasses many aspects, such as identification, availability, privacy and accountability. 
This paper considers the security of cellular networks serving IoT devices and proposes the use of an unmanned aerial system (UAS) to act as a network support node to improve the data confidentiality or secrecy of the data. 
More precisely, we analyze the secrecy rate of using a moving UAS serving a mobile IoT device. 

The UAS or unmanned aerial vehicle (UAV) has recently gained attention for supporting non-military applications~\cite{jaber2017}. 
A future UAS can be seen as an IoT device that collects sensor data, for example. 
The UAS can also carry several IoT devices, or serve as the collector of sensor data.  
In the latter case, 
an UAS node can be used as an access point for IoT sensors to upload their data while the UAS is close enough for reliable and low power transmission, a need for most IoT devices.

Communications systems for UAVs or with UAVs have been analyzed by the research community and industry \cite{8618602}. 3GPP, in particular, is evaluating UAV use cases and developing specifications for supporting UAVs with 5G technology. 
This paper is agnostic to the communications protocol and provides early results on UAV-assisted networking for the IoT, considering mobility. 
The rest of the paper is organized as follows: 
Section II discusses the fundamental wireless security principles and how they apply to this context. 
Section III introduces the system models for the air-to-ground (A2G) and ground-to-ground (G2G) channels, for the mobility and for the secrecy rate. 
Section IV provides a numerical analysis of the effect of using aerial relays for increasing the channel secrecy rate against eavesdropping. 
Our results show that the mobility of UAVs and the properties of channels can be leveraged to increase the secrecy rate. 
Section V concludes the paper. 

%% file: include/requirements.tex
There are different types of attacks that can 
compromise wireless communications systems. 
The attack can come from an UAS and can affect the safe operation of aerial vehicles, which are characterized by high maneuverability, dynamic changes in mobility patterns and often dominant line of sight (LoS) radio links.

\subsection{Identification, Authenticity, and Integrity}

Any cellular network user needs to be identified and authorized by the network to access its services. 
In addition, the network needs to be authenticated by the users so that mutual trust can be established.
Man in the middle attacks affect the integrity of messages by capturing the transmitted data and forwarding a manipulated version to the destination~\cite{10}

\subsection{Availability}

Capacity is a common metric used for evaluating modern wireless communications systems. 
Malicious users can exploit the fact that radio resources are limited and can leverage knowledge about how they are managed by the network.
If, for example, malicious users create fake service requests or participate in broadcasting fake messages, the radio frequency (RF) spectrum will become congested and the communications services less reliable and potentially less available~\cite{7}. 
Mechanisms that allow recovering corrupted or lost messages are usually based on retransmission and additional processing. 
This adds to the RF congestion, increases latency and power consumption, among others. 
GPS spoofing attacks typically provide false location coordinates and, possibly, time offsets which can then lead to RF interference and service unavailability. 

\subsection{Confidentiality and Privacy}
Eavesdroppers can capture packets or monitor user/network activity to compromise confidentiality and privacy~\cite{9}. 
The identities, positions, actions and trajectories of mobile users/IoT devices therefore need to be securely transmitted to avoid unauthorized tracking of devices and exploitation of data. 
For most cases, it is important that  devices regularly report their presence and control or sensor data. 
This increases the attack surface and poses security challenges, especially for the IoT, where resources are limited. 

\subsection{Non-Repudiation and Accountability}

A malfunctioning or malicious device 
can significantly compromise the radio access network performance. 
For example, a device that is not adhering to 
the time advance commands sent by the serving base station, can create significant interference from its transmission. 
Many outdoor devices use GPS signals as the synchronization source, but can use infrastructure nodes, where available, or other sources. If no external source exists, 
frequency and timing drifts will occur that add up over time. This can cause significant interference and system malfunctioning. 
The system needs to monitor such 
transmissions, or unauthorized transmissions, in general, and make the corresponding devices accountable before they cause major damage to the network operation.\\

%% file: include/problem.tex
\subsection{Problem Formulation}

The eavesdropping attack is the ability of a malicious node to intercept exchanged packets between legitimate nodes or fixed infrastructure. This type of attack can be very harmful for IoT networks because of the anonymous leakage of confidential information without legitimate transmitter/receiver knowledge. 
To this end, the aerial relay has been proposed as a mitigation technique to minimize the eavesdropper's wiretap link rate. 
The objective of this paper is to 
illustrate how UAV path planning and speed of the aerial relay can influence the secrecy rate of a mobile IoT device in the presence of an eavesdropper. 

\begin{figure}[t]
    \centering
    \includegraphics[width=3.5in, height=3.3in]{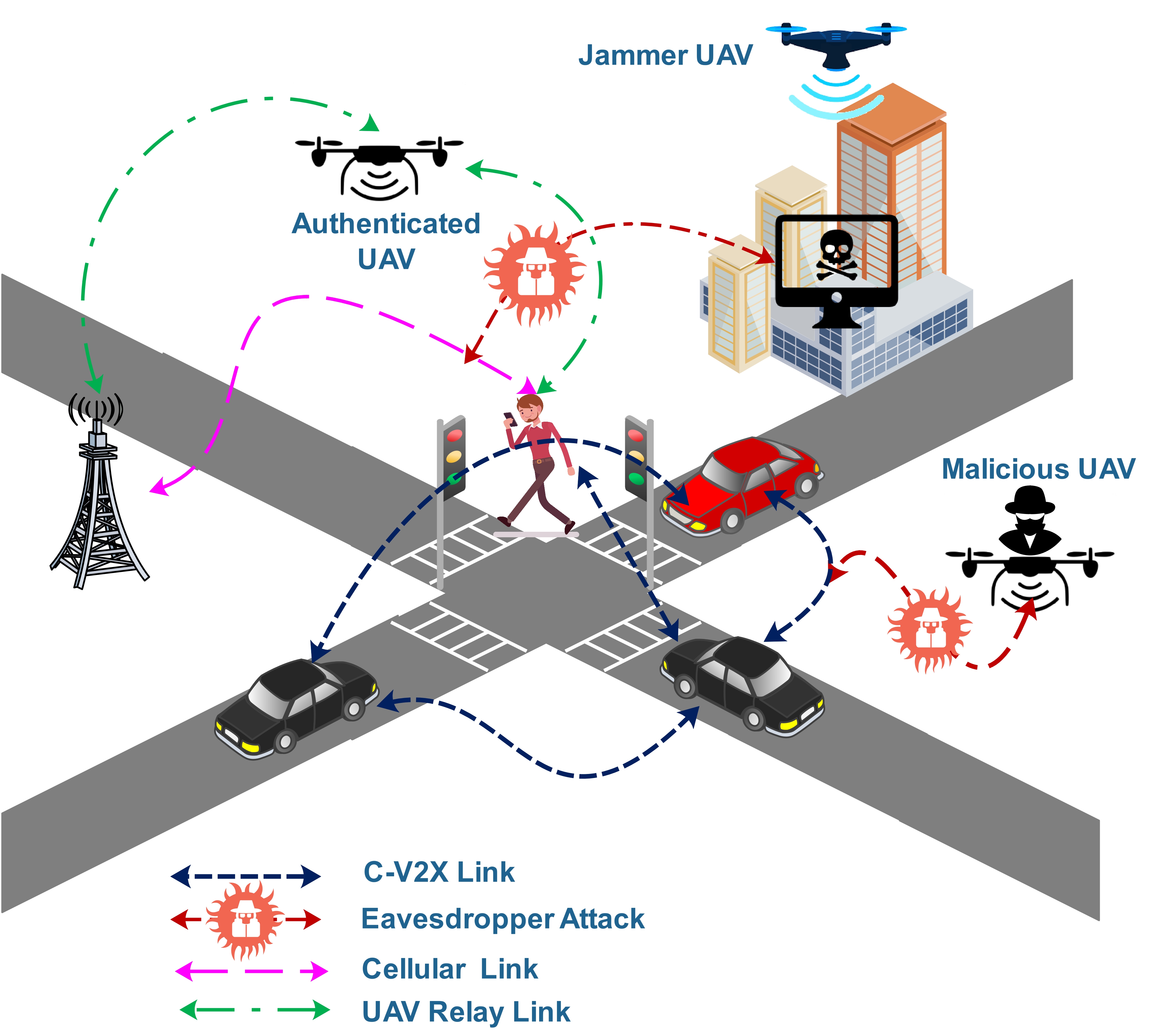}
    \caption{UAVs as authenticated and malicious nodes in a 5G cellular network.}
    \label{fig:Figure1}
\end{figure}

Figure~\ref{fig:Figure1} illustrates a scenario where an authenticated UAV is relaying messages between a cellular base station and 
ground IoT device, while a ground eavesdropper is attacking the direct link between the user equipment (UE) and the base station. 

\subsection{Related Work}
Early research has shown how UAVs can extend cellular networks and be used to improve the security of terrestrial networks~\cite{wu2019safeguarding,shang,wang}. 
References~\cite{wu2019safeguarding} and~\cite{shang} propose using UAVs as jammers to enhance the physical layer security of cellular networks. They investigate the effects of having a strong LoS jamming link between the UAV and a single or multiple eavesdroppers. Reference~\cite{wang} discusses the use of UAVs as mobile relays to improve the results of static relays for static ground users. The authors use the difference-of-concave (DC) program to solve the secrecy rate maximization problem and find that each DC iteration yields a closed-form solution.  

This paper extends those initial results~\cite{wu2019safeguarding,shang,wang} and considers the UAV dynamics to gain new insights on how UAVs can be effectively used to 
increase the secrecy of the channel. We consider the mobility parameters of the UAV and study the secrecy rate for mobile IoT ground devices. We compare the proposed UAV relaying approach to a base station handover. 

%% file: include/system.tex
% ...

% ...



% ...
\subsection{Air-to-Ground Channel Model}
\label{sec:A2G}
For a ground receiver, the received signals from a UAV features the LoS signal, non-LoS (NLoS) signals and multiple reflected components which cause multipath fading~\cite{6863654}.
The resulting A2G mean pathloss can be calculated as
\begin{equation}
    P{L_\varsigma }(d,{h_t},{h_r}) = FSPL\left( d \right) + {\eta _\varsigma }\left( {{h_t},{h_r}} \right) ,
\end{equation}
where $\varsigma  = \left\{ {LOS,NLOS} \right\}$, $d$ is the communication distance, $h_t$ and $h_r$ denote the heights of the transmitter (Tx) and receiver (Rx), $FSPL\left( d \right) = 20{\log _{10}}\left( {\frac{{4\pi fd}}{c}} \right)$ is the free space pathloss model, and ${\eta _\varsigma }\left( {{h_t},{h_r}} \right)$ is the mean value of the excessive pathloss.
The parameters $f$ and $c$ stand for the carrier frequency and the speed of light.

According to field test measurements~\cite{6863654}, the occurrence probability of a LOS link ${\rho _{LoS}}(\theta )$ between a UAV transmitter and a ground receiver is given by
\begin{equation}
\begin{aligned}
    &{\rho _{LoS}}(\theta ) = \frac{1}{{1 + C\exp \left[ { - B\left( {\theta  - C} \right)} \right]}} , \\
    &\theta  = \frac{{180}}{\pi }\arctan \left( {\frac{{{h_t}}}{r}} \right) ,
\end{aligned}
\end{equation}
where $C$ and $B$ are constant values that depend on the communications environment, e.g. rural, urban, or dense urban. 
Parameter $h_t$ represents the height of the UAV transmitter and $r$ the horizontal distance between the UAV and the ground receiver.
Note that the LoS and NLoS probabilities are related as ${\rho _{NLoS}}(\theta ) = 1 - {\rho _{LoS}}(\theta )$.

The spatial expectation of pathloss is given by
\begin{equation}
\begin{aligned}
  PL(\theta ,d,{h_t},{h_r}) = \sum\limits_\varsigma  {{\rho _\varsigma }(\theta )P{L_\varsigma }(d,{h_t},{h_r})} \label{PL} .
\end{aligned}
\end{equation}
The path loss is higher for a NLoS than for a LoS link because of shadowing and indirect signal paths.
We use the parameter $\eta  = \frac{{{\eta _{LOS}}}}{{{\eta _{NLOS}}}} $ 
to denote the ratio of excessive attenuation factor for NLoS links compared to LOS links.

The Nakagami-m distribution is used to describe the small scale fading $g$ in A2G channels. 

\subsection{Ground-to-Ground Channel Model}
\label{sec:A2A}
For ground communications, we consider both the distance dependent large-scale fading and small-scale fading.
The signal transmitted from the ground transmitter is attenuated with a coefficient ${f_G} = g{d^{ - \alpha }}$, where $g$ is the channel power gain with exponential distribution, $d$ is the communication link distance between the ground transmitter and receiver, and $\alpha$ is the path-loss exponent.

The signal-to-noise ratio (SNR) of ground communications is given by
\begin{equation}
\begin{aligned}
  SN{R_G} = \frac{{{P_T}{f_G}}}{{B{n_0}}}  ,
\end{aligned}
\end{equation}
where $P_T$ indicates the transmit power of the ground transmitter, $B$ is the channel bandwidth and $n_0$ is the noise spectral density.

\subsection{Mobility Model}
\label{sec:mobility}
In this paper, we consider a mobile IoT device.
We examine the secrecy rate of this mobile IoT device with a static eavesdropper.
Without loss of generality, the mobility of the ground user and the aerial relay will be over the x-axis with fixed y-position for all nodes (Tx, Rx, and relay). First we define two parameters that will be used for deriving our mobility model:\\
\begin{itemize}
\item \textbf{The distance step ($dx$)} is a constant step size that corresponds to the granularity of movement. \\
\item \textbf{The speed rate ($SR$)} is a ratio that defines how fast the UAV flies with respect to the mobile IoT device. 
When $SR = 1$, the UAV will maintain the same speed as the moving IoT device and closely follow it (same x coordinates). When $SR > 1$, the UAV will be in front of the ground IoT receiver. 
When $SR < 1$, the UAV will lag the IoT receiver. 
\end{itemize}{}

The following model is used to define the mobility of the ground IoT device and the UAV relay:
\begin{equation}
    IoT_{X_{t+1}} = IoT_{X_{t}} + dx,
\end{equation}{}
\begin{equation}
    UAV_{X_{t+1}} = SR * IoT_{X_{t+1}},
\end{equation}{}where $IoT_{X_{t+1}}$ and $UAV_{X_{t+1}}$ are the next x-positions of the ground IoT device and aerial relay in the next time step. 
$UAV_{X_{t+1}}$ is calculated as a function of the position of IoT device and the speed rate. 

\subsection{Secrecy Rate}
In this subsection, we present the secrecy rate of the ground legitimate user where an authenticated UAV serves as an aerial relay.
The UAV forwards the user information. The secrecy rate ${R_{Sec}}$ of the ground legitimate user is then calculated as the difference of the legitimate information rate ${R_L}$ and the interception rate ${R_I}$, i.e., ${R_{Sec}} = \max \left( {{R_L} - {R_I},0} \right)$.
More specifically, we have
\begin{equation}
 \begin{aligned}
  {R_L} = \max \left( {{R^{L}_{T}},{R^{L}_{R}}} \right), \label{III-C-1}
 \end{aligned}
\end{equation}
\begin{equation}
 \begin{aligned}
  {R_I} = \max \left( {{R^{I}_{T}},{R^{I}_{R}}} \right) , \label{III-C-2}
 \end{aligned}
\end{equation}
where $R^{L}_{T}$ is the data rate from the legitimate transmitter to the 
legitimate receiver, $R^{L}_{R}$ is the data rate from the legitimate aerial relay to the legitimate receiver, $R^{I}_{T}$ is the data rate from the legitimate transmitter to the eavesdropper, and $R^{I}_{R}$ is the data rate from the legitimate aerial relay to the eavesdropper.
Considering a UAV that serves as an aerial relay, we split a time-slot into two parts.
One is for the UAV receiving information from the ground transmitter, and the other is for UAV transmitting data to the ground receiver.

More specifically, we have
\begin{equation}
 \begin{aligned}
  R_T^L = B{\log _2}\left( {1 + \frac{{g{{\left\| {{x_{BS}} - {x_{UE}}} \right\|}^{ - \alpha }}}}{{B{n_0}}}} \right) ,
 \end{aligned}
\end{equation}
\begin{equation}
 \begin{aligned}
  R_R^L = &B{\log _2}\left( {1 + {{1 \mathord{\left/
 {\vphantom {1 {B{n_0}}}} \right.
 \kern-\nulldelimiterspace} {B{n_0}}}} } \right. \hfill \\
  &\cdot \left. {PL({\theta _{UAV - UE}},\left\| {{x_{UAV}} - {x_{UE}}} \right\|,{h_{UAV}},{h_{UE}})} \right) ,
 \end{aligned}
\end{equation}
\begin{equation}
 \begin{aligned}
  R_T^I = B{\log _2}\left( {1 + \frac{{g{{\left\| {{x_{BS}} - {x_{EAV}}} \right\|}^{ - \alpha }}}}{{B{n_0}}}} \right) ,
 \end{aligned}
\end{equation}
\begin{equation}
 \begin{aligned}
  R_R^I = &B{\log _2}\left( {1 + {{1 \mathord{\left/
 {\vphantom {1 {B{n_0}}}} \right.
 \kern-\nulldelimiterspace} {B{n_0}}}} } \right. \hfill \\
  &\cdot \left. {PL({\theta _{UAV - EAV}},\left\| {{x_{UAV}} - {x_{EAV}}} \right\|,{h_{UAV}},{h_{EAV}})} \right) ,
 \end{aligned}
\end{equation}
where $\left\| {a - b} \right\|$ denotes the distance between $a$ and $b$, $PL( \cdot )$ is the spatial expectation pathloss of A2G channel which is given by~(\ref{PL}).
Therefore, the secrecy rate of the ground IoT device is given in (\ref{R_sec}) at the top of next page.
\begin{figure*}[h]
\setlength{\abovecaptionskip}{-0.5cm}
\setlength{\belowcaptionskip}{-0.5cm}
\normalsize
\begin{equation}
\begin{split}
  {R_{\operatorname{Sec} }} = \max \left( {B{{\log }_2}\left( {\frac{{B{n_0} + \max \left( {g{{\left\| {{x_{BS}} - {x_{EAV}}} \right\|}^{ - \alpha }},PL({\theta _{UAV - EAV}},\left\| {{x_{UAV}} - {x_{EAV}}} \right\|,{h_{UAV}},{h_{EAV}})} \right)}}{{B{n_0} + \max \left( {g{{\left\| {{x_{BS}} - {x_{UE}}} \right\|}^{ - \alpha }},PL({\theta _{UAV - UE}},\left\| {{x_{UAV}} - {x_{UE}}} \right\|,{h_{UAV}},{h_{UE}})} \right)}}} \right),0} \right) .   \label{R_sec}
\end{split}
\end{equation}
\hrulefill
\end{figure*}

%% file: include/results.tex
\begin{figure}[h]
    \centering
    \includegraphics[width=3.5 in, height=3.3in]{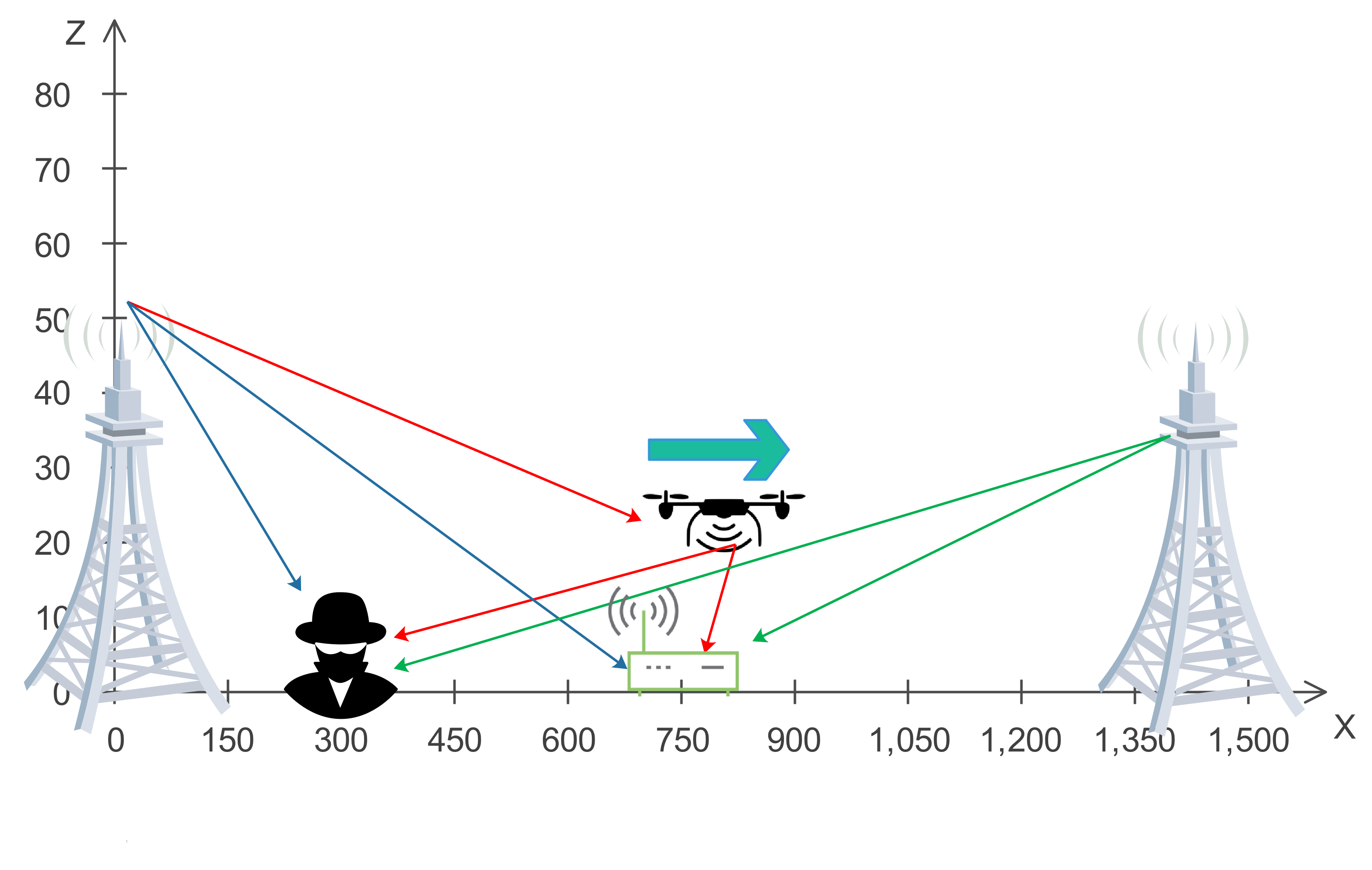}
    \caption{Simulation scenario.}
    \label{fig:Figure4}
\end{figure}

We simulate two use cases. The first use case compares the efficiency of the aerial relay with a traditional handover procedure between two base stations as a mitigation technique against eavesdropping. This is illustrated in Fig.~\ref{fig:Figure4}. 
The second use case investigates the effect of the aerial relay on the secrecy rate as a function of the position and speed of the relay with respect to the position of the IoT device. 


\subsection{Mobile IoT and UAV Relay Use Cases}
The overarching scenario is that of a base station serving IoT devices. We consider the downlink data transmission, which can be a combination of user and control data or configuration signals. 
This is applicable to many applications, such as private content downloading, or software upgrades.
For the sake of illustration and without loss of generality, we consider a single IoT device that may represent several nearby devices. 

The IoT device and the UAV relay, when used, are moving on a straight line. They may, for example, follow a long straight road. 
The IoT device and the UAV relay move at some speeds, which are not necessarily constant nor identical. 
Two base stations and one eavesdropper are considered at fixed locations on the same road. 
When the UAV relay is used, the base station uses beamforming to the UAV and the fronthaul channel (from the base station to the UAV) is assumed to be at full capacity and orthogonal to the access channel (UAV to ground devices). 

The UAV uses an omnidirectional antenna. 
The path loss exponent is 4 and the Rayleigh distribution is used to model the small-scale fading of ground links. 
The remaining simulation parameters are provided in Table~\ref{tab:my-table1}.

This scenario can be scaled by adding additional IoT devices, base stations, UAVs and eavesdroppers. When more nodes are used, their locations and trajectories need to be defined in the three dimensional space.

\subsection{UAV Relay vs. Base Station Handover}

\subsubsection{Simulation Scenario}

The ground IoT device  starts moving from the initial position following the mobility model introduced in Section~\ref{sec:mobility}. 
The aerial relay starts flying and following the same trajectory as the IoT device. 
The IoT device is connected to the primary base station from the start. 
When the secrecy rate of the channel between the primary base station and the IoT receiver drops, 
handover to another base station is considered and the handover initiated when the secrecy rate provided by the new association is found to be higher. 

\begin{figure}[h]
    \centering
    \includegraphics[width=3.7in, height=2.7in]{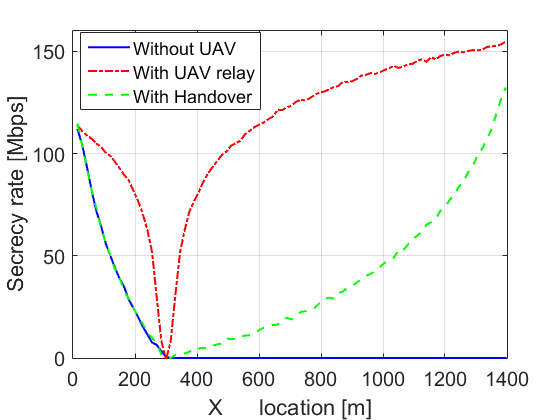}
    \caption{Secrecy rate without a relay, with a UAV relay, and with base station handover.}
    \label{fig:Figure2}
\end{figure}

\label{sec:relay}
\begin{table}[]
\fontsize{4}{6}\selectfont
\centering
\caption{Simulation parameters.}
\label{tab:my-table1}
\resizebox{0.75\columnwidth}{!}
{
\begin{tabular}{|c|c|}
\hline
\begin{tabular}[c]{@{}c@{}}\textbf{Simulation  parameter}\end{tabular}     & \textbf{Value}                              \\ \hline
\begin{tabular}[c]{@{}c@{}}Primary base \\ station position\end{tabular}         & (0,0,50)                           \\ \hline
\begin{tabular}[c]{@{}c@{}}Secondary base \\ station position\end{tabular}         & (1400,0,50)                         \\ \hline
\begin{tabular}[c]{@{}c@{}}RX(IoT)\\ initial position\end{tabular} & (0,0,0)                            \\ \hline
\begin{tabular}[c]{@{}c@{}}Eavesdropper\\ position\end{tabular}    & (300,0,0)                          \\ \hline
\begin{tabular}[c]{@{}c@{}}UAV relay\\ position\end{tabular}       & (0, 0, 20) \\ \hline
dx                                                                 & 15 m                               \\ \hline
SR                                                         & 1, 2, 0.5,  and 0.75                   \\ \hline
Base station power                                                           & 0.1 watt                           \\ \hline
UAV power                                                          & 0.01 watt                              \\ \hline
Noise variance                                                     & 10$^{\text{-12} }$       \\ \hline
Bandwidth                                                          & 10 MHz                              \\ \hline
\end{tabular}%
}
\end{table}

\begin{figure*}[h]
    \centering
    \includegraphics[width=6.2in, height=2.7in]{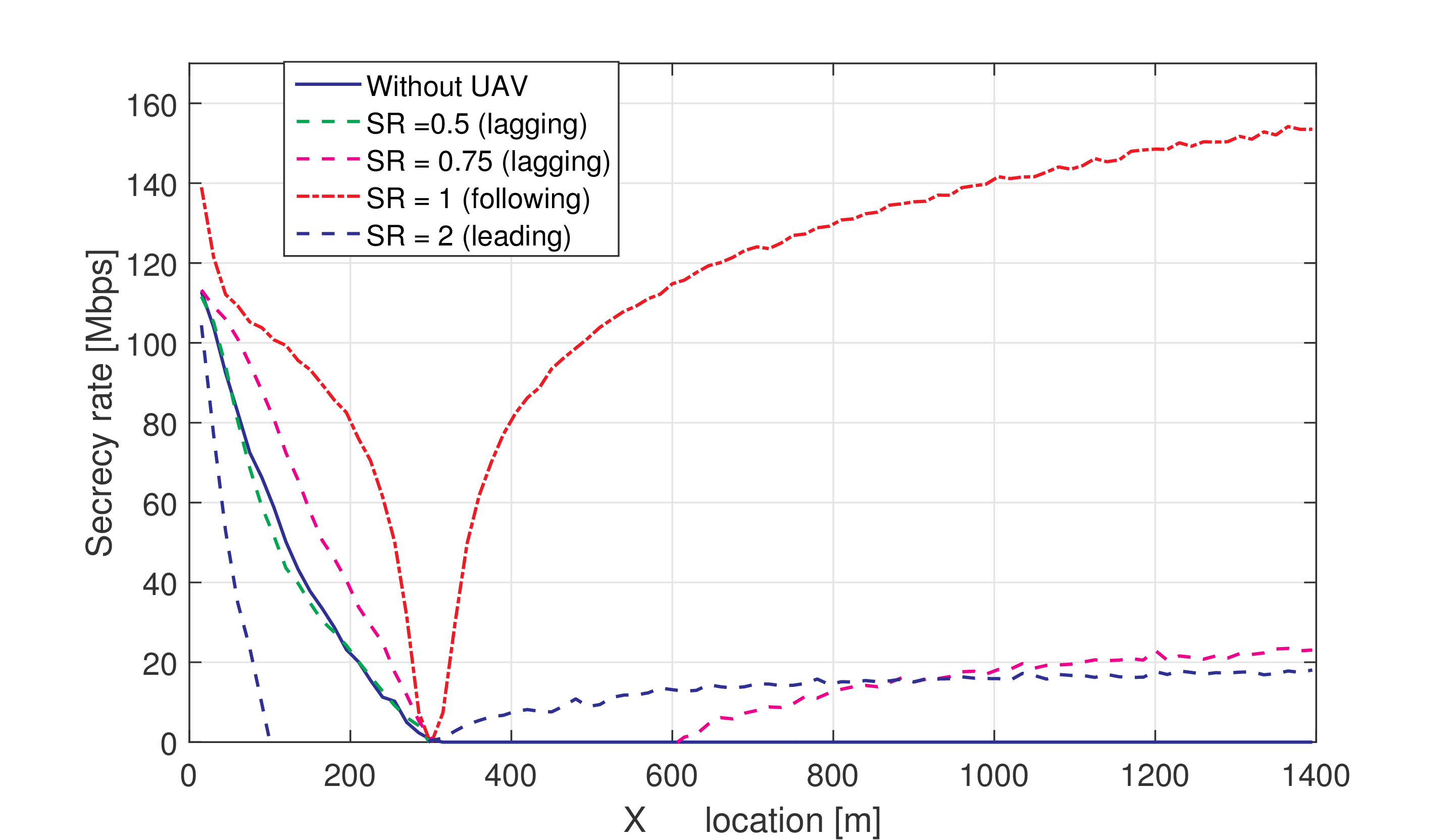}
    \caption{Secrecy rate comparison between different aerial relay tracking paths with respect to the mobile IoT device.}
    \label{fig:Figure3}
\end{figure*}
\subsubsection{Results}
The results are illustrated in Fig.~\ref{fig:Figure2}, which compares the secrecy rates of the three strategies: normal operation, base station handover, and UAV relaying.  
The secrecy rate for normal operation drops gradually until reaching 0 at the location of the eavesdropper and not recovering when the IoT device passes it. 
Using a second base station and initiating an early handover allows the secrecy rate to recover, but slowly because of the larger distance to the second base station. 
With the UAV relay, the secrecy rate is maintained at high levels over a wide range. The rate drops steeply to zero when the IoT device and the eavesdropper meet, but it also recovers quickly. 
It is below 50 Mbps only while the IoT device is within 50 m distance to the eavesdropper. 



We conclude that the use of an aerial relay as a mitigation solution for eavesdropping is more efficient than using traditional mitigation techniques, such as handover. 

\subsection{UAV Relay Trajectory}
\label{sec:trajectory}


\subsubsection{Simulation Scenario}
The second case study analyzes the effect of the UAV position with respect to the mobile ground receiver on the secrecy rate, where 
both are mobile and move in concert.
We analyze the effect of different UAV speeds with respect to the IoT device. 
We implement three approaches of UAV mobility, leveraging the models introduced in Section~\ref{sec:mobility}: leading (SR = 2), lagging (SR = 0.5, 0.75), and following (SR = 1). 
Since the UAV and IoT device start at the same point, the leading UAV flies ahead and away from the IoT device, the lagging UAV behind it and the following right above it. 
The IoT device and aerial relay move away from the primary base station and pass the eavesdropper at the fixed location of 300 m. 
Only one base station is involved in this scenario. 


\subsubsection{Results}
Fig.~\ref{fig:Figure3} shows the simulation results. 

we observe that the UAV that closely follows the IoT device achieves the best secrecy rate overall, which is mostly above 50 Mbps, whereas for the leading or lagging UAV the secrecy rate is mostly below 50 Mbps. 
Using a small lagging factor between UAV relay and moving ground legitimate user before reaching the position of the eavesdropper is providing a higher secrecy rate than the leading case. However, the leading case exhibits a rapid recovery of the secrecy rate after passing the eavesdropper position. Also, as the gap between UAV relay and the moving ground user widens, for both lagging or leading UAVs, the performance of the secrecy rate declines. This is result of lower legitimate rate between a ground user and UAV because of lower SNR. 
A leading or lagging UAV may still be a better option for more complex terrains and locations of the transmitter, receiver and eavesdropper. 
This has to do with the terrain surrounding the nodes and, hence, radio channel. 
In general, the UAV should position itself such that it provides a high data rate to the IoT device and a low data rate to the eavesdropper. 

%% file: include/conclusions.tex
This paper considers the mobile IoT context 
and analyzes how UAVs can be deployed to improve the secrecy rate of terrestrial IoT devices being served by cellular networks. 
IoT devices may be carried by vehicles, but many devices will likely have lower protection mechanisms against security exploits because of cost or other resource constraints. 
An eavesdropper may thus be able to sniff and decode the transmitted data. 
Besides many cellular control channels are sent in the clear and the eavesdropper may gain insights simply from RF monitoring of control data. 
Under these premises, we have modeled and numerically analyzed the secrecy rate of the wireless channel in the context of mobile IoT with eavesdropping. 
The simulation results provide new insights on the effectiveness of physical-layer security against eavesdropping using UAV relays. 
The results show that a single aerial relay considerably increases the secrecy rate and largely outperforms the traditional approach of user association to another base station. 
These results are promising and extend early works that explore 
the use of UAVs to improve the security of cellular communications. 

Further research is needed for fundamentally solidifying the proposed concepts as well as for integrating them into commercial communications systems. 
Experimental research will be critical for this and requires large-scale test facilities that enable 3D mobility experiments. 
The Aerial Experimentation and Research Platform for Advanced Wireless (AERPAW) will provide such a testbed where modern communications systems and UAVs will be operated in real operating environments~\cite{aerpaw_vtc19}. 
AERPAW will allow evaluating 
trajectory and UAV speed optimization algorithms, as well as exploring coordinated mechanisms between terrestrial and aerial networking nodes and their different modes of operation. 